\documentclass[usenatbib]{mn2e}
\usepackage{graphicx}
\usepackage{amsmath,amssymb}
\title[How Well Do We Know The Halo Mass Function?]
	{How Well Do We Know The Halo Mass Function?}
\author[S.~G.~Murray, C.~Power \& A.~S.~G.~Robotham]
	{S.~G.~Murray$^{1,2}$\thanks{steven.murray@uwa.edu.au}, 
	 C.~Power$^{1,2}$ \& A.~S.~G.~Robotham$^{1}$\\
	$^1$ ICRAR, University of Western Australia, 35 Stirling Highway,
	Crawley, Western Australia 6009, Australia\\
        $^2$ ARC Centre of Excellence for All-Sky Astrophysics (CAASTRO)}

\begin{document}

\maketitle

\begin{abstract} 
The parameters governing the standard $\Lambda$ Cold Dark Matter cosmological
model have been constrained with unprecedented accuracy by precise measurements
of the cosmic microwave background by the Wilkinson Microwave Anisotropy Probe 
(WMAP) and Planck satellites. Each new data release has refined further our 
knowledge of quantities -- such as the matter density parameter 
$\Omega_{\rm M}$ -- that are imprinted on the dark matter halo mass function 
(HMF), a powerful probe of dark matter and dark energy models. In this 
letter we trace how changes in the cosmological parameters over the last 
decade have influenced uncertainty in our knowledge of the HMF. We show that 
this uncertainty has reduced significantly since the $3^{\rm rd}$ WMAP data 
release, but the rate of this reduction is slowing. This is limited by 
uncertainty in the normalisation $\sigma_8$, whose influence is most 
pronounced at the high mass end of the mass function. Interestingly, we find 
that the accuracy with which we can constrain the HMF in terms of the 
cosmological parameters has now reached the point at which it is comparable 
to the scatter in HMF fitting functions. This suggests that the power of the 
HMF as a precision probe of dark matter and dark energy hinges on more 
accurate determination of the theoretical HMF. Finally, we assess prospects 
of using the HMF to differentiate between Cold and Warm Dark Matter models 
based on ongoing improvements in measurements of $\Omega_{\rm M}$, and we comment
briefly on optimal survey strategies for constraining dark matter and dark energy
models using the HMF.

\end{abstract}

\section{Introduction}
\label{sec:intro}
The halo mass function (hereafter HMF), which encodes the comoving number 
density of dark matter haloes in the Universe at a given epoch as a function of 
their mass, is a powerful probe of cosmology, dark matter and dark energy 
\citep{Press1974,Jenkins2001,Tinker2008,Vikhlinin2009}. For example, the 
amplitude of the HMF on the scale of galaxy clusters at the present epoch
may be used to deduce limits on the combination of the power spectrum normalisation
$\sigma_8$ and the matter density parameter $\Omega_{\rm M}$ \citep{Vikhlinin2009,Allen2011}. 
Similarly, the evolution of this amplitude over cosmic time may be used to
characterise the dark energy equation of state $w_0$ \citep{Vikhlinin2009,Allen2011}. 

Over the last decade, precise measurements of the cosmic microwave 
background (CMB) by the Wilkinson Microwave Anisotropy Probe
\citep[hereafter WMAP; cf.][]{Spergel2003} and Planck \citep{PlanckCollaboration2013} 
satellites have led to increasingly accurate estimates of key cosmological parameters 
such as $\Omega_{\rm M}$, by factors of $\sim$4 in most cases. It is
therefore interesting to ask how our knowledge of the HMF has evolved over the same 
period. 

In this Letter, we estimate the uncertainty in the HMF assuming
``best-bet'' WMAP and Planck cosmological parameters, and we determine the 
independent parameters that are the primary sources of this uncertainty. 
This is of crucial importance because quantifying the significance of any 
deviation between an observationally derived HMF and one predicted within 
the standard cosmological framework is difficult without understanding the 
framework's instrinsic uncertainties. Specifically, we determine
uncertainties in the predicted HMF for a suite of flat $\Lambda$
Cold Dark Matter (hereafter $\Lambda$CDM) cosmologies, adopting a range of HMF fitting 
functions drawn from the literature. We present the 68\% error on the amplitude 
and slope of the HMF, given the reported errors on a number of input 
parameters; a comparison of errors due to uncertainty in cosmology with 
errors in the chosen fitting function; and an analysis of the parameters 
that provide the primary sources of uncertainty. Finally, we consider the 
sensitivity of the HMF to the assumed dark matter model by exploring the 
range of Warm Dark Matter (WDM) particle masses for which the HMF can be used to
differentiate between $\Lambda$CDM and its $\Lambda$WDM alternatives.

\section{Methodology}
\label{sec:method}
We calculate the HMF using the formalism of \cite{Press1974} and \cite{Bond1991}. 
This defines the HMF as the differential number density of haloes in 
logarithmic mass bins,
\begin{equation}
  \label{eq:hmf}
  \frac{dn}{d\log M} = \frac{\rho_0}{M} f(\sigma) \left|\frac{d\ln\sigma}{d\ln M}\right|,
\end{equation}
where $\rho_0$ is the mean matter density of the universe; $\sigma$ 
is the mass variance at mass scale $M$; and the function $f(\sigma)$ 
differentiates between fitting functions. 

Many forms for $f(\sigma)$ have been proposed in the literature. The original 
form proposed by \cite{Press1974} is the only one derived completely analytically, 
but makes the simplifying assumption of collisionless spherical collapse. 
\cite{Sheth2001} proposed a form motivated by the more general assumption of ellipsoidal 
collapse in which three parameters were set by fitting to simulation 
data. To date, no other analytical form for $f(\sigma)$ has been derived; 
instead, its form is empirically derived by fitting to the abundance of haloes measured 
by halo finders in cosmological simulations. In this Letter we adopt many of these
empirical fitting functions from the literature in addition to the forms of 
\cite{Press1974} and \cite{Sheth2001}. For clarity, we focus on the form of Sheth-Tormen 
unless otherwise stated; we note that results are qualitatively similar for all of the 
fitting functions we have considered.

We calculate HMFs using the \verb|hmf| code
(cf. {\small github.com/steven-murray/hmf}), which is 
the backend of the \verb|HMFcalc| 
web-application (cf. {\small hmf.icrar.org}); further details can be found in
Murray, Power \& Robotham (In Prep.). Our code uses \verb|CAMB| 
\citep[cf. {\small camb.info}; details in ][]{Lewis2000} to produce a transfer function for 
a given input cosmology, and interpolates and integrates this function to compute the 
mass variance. Importantly, our code is optimized to quickly and easily update
parameters of the calculation, which allows us to generate the large number of 
realisations necessary for this study. 

We use data spanning the last 7 years of cosmological parameters derived
from the CMB -- from WMAP3 \citep{Spergel2007}, WMAP5 \citep{Komatsu2009}, WMAP9 
\citep{Hinshaw2012a} and Planck \citep{PlanckCollaboration2013}. We choose parameter
sets that derive from isolated fits to the CMB, not using any extra data (such as BAOs 
or lensing) to ensure consistency across samples. In all cases we assume a flat 
$\Lambda$CDM geometry, consistent with the theoretical base model of the Planck results, 
and consider five parameters -- the baryon and CDM densities combined with the hubble 
constant $\Omega_{\rm b}h^2$ and $\Omega_{\rm c}h^2$, the spectral index $n_s$, the Hubble 
parameter $H_0$ and the normalization $\sigma_8$. We constrain $H_0$ by adopting Eq. 11 of 
\citet{PlanckCollaboration2013}; this is possible because
we can calculate directly the angular size of the sound horizon at last scattering, to better than 0.3\% accuracy.

In order to sample from the parameter distributions from each base cosmology, 
we use available Monte Carlo Markov chains (MCMC), randomly choosing 5000 realizations 
from each chain. This allows for robust sampling of the covariance between 
each parameter. We calculate the HMF for each realization, finding the resulting 
68\% uncertainty about the median for both the amplitude and slope. Note that we have 
checked our results for consistency with more basic approaches based on (i) variance in 
the parameters, in which parameter uncertainties are assumed to be uncoupled, and (ii) 
covariance between the parameters, in which parameter uncertainties are coupled but assumed 
to be Gaussian-distributed. Averaging over masses, we find that the maximum difference 
between the variance and MCMC approaches is 6\% (for WMAP3), while 
the maximum difference between the covariance and MCMC approaches is 2\% (for Planck).
If we do not average over masses, we find that the variance approach underestimates 
the uncertainty in the HMF by $\sim 25\% (10\%)$ for all WMAP (Planck) datasets at high 
masses ($\sim 10^{15} h^{-1} {\rm M}_{\odot}$), whereas it is $\sim 2\%$ at most if we 
adopt the covariance approach.

The amplitude uncertainty range is simply the 16$^{th}$ -- $84^{th}$ 
quantile of the value of the HMF. Calculation of this quantity 
eliminates any information associated with the gradient of the mass 
function, which may be an important aspect in constraining cosmology. 
Therefore we also calculate the slope, as the arc-tangent of the discrete 
gradient (via the method of central differences) at each mass bin. For 
this analysis we use a mass bin width of 0.05 dex.

Note that we perform our analysis assuming a ``universal" form for 
the mass function, in which changes in cosmology are captured by changes 
in the mass variance $\sigma$. Recent work suggests that this assumption 
does not hold in detail and can lead to $\sim$10\% uncertainties 
\citep[cf.][]{Tinker2008,Bhattacharya2011}. We do not explicitly account
for this source of uncertainty in our analysis and we suggest that the reader
allow for an additional 10\% uncertainty in our quoted results to account
for this.


\section{Results}
\label{sec:results}
\begin{figure}
  \centering
  \includegraphics[width=\linewidth, trim=0.5cm 0cm 0.3cm 0cm, clip=False]{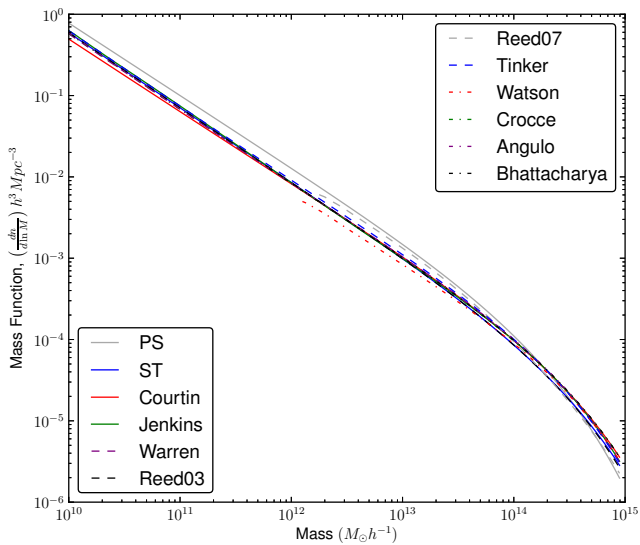}
  \caption{The HMF for several fitting functions drawn from the literature, 
    using a Planck \citep{PlanckCollaboration2013} cosmology; see text for further
    details.}
  \label{fig:hmf}
\end{figure}

\paragraph*{Effect of Choice of Fitting Function.}
In Figure \ref{fig:hmf} we show the variety of HMF fitting functions drawn
from the literature -- \cite{Press1974,Sheth2001,Jenkins2001,Reed2003,Warren2006,Reed2007,Tinker2008,Crocce2010,Courtin2010,Bhattacharya2011,Angulo2012} and \cite{Watson2012} -- 
assuming a Planck cosmology.
The scatter between fitting functions is noticeable and has been remarked upon already
in the comprehensive comparison of halo-finders presented in 
\cite{Knebe2011}. Given this scatter, we wish to investigate two effects;
(i) whether or not the fitting functions behave similarly under changes of 
cosmology, and (ii) whether or not the scatter in the fitting functions is 
the dominant source of uncertainty in the HMF for contemporary cosmological 
parameter sets. That is, whether the variation between fitting functions for 
a cosmology with no variance is greater than the variation between HMF's of a 
single fitting function for a cosmology with non-zero variance. 

\begin{figure}
  \centering
  \includegraphics[width=\linewidth, trim=2cm 0cm 2.5cm 0cm, clip=True]{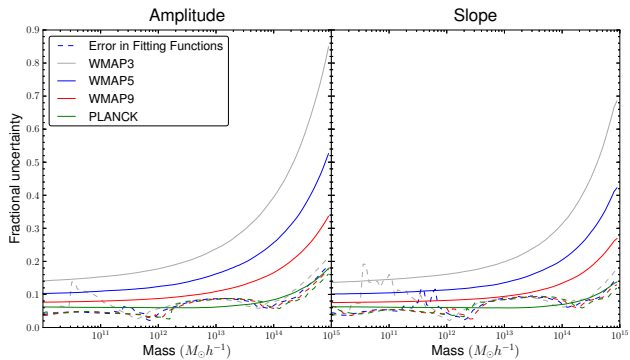}
  \caption{Errors on the amplitude (left) and slope (right) of the HMF 
    for a range of cosmological parameter sets, as a fraction of the HMF. 
    Solid lines indicate intrinsic parameter error, while dashed lines 
    indicate error due to the choice of fitting function.}
  \label{fig:error}
\end{figure}

In Figure \ref{fig:error} we show the fractional error intrinsic to each 
parameter set; the Sheth-Tormen fitting function is shown by the solid lines, 
while the fractional error attributable to the choice of fitting function, 
estimated at the median value of the HMF for the given cosmology, is
shown by the dashed lines. It is noticeable that the variance in earlier
parameter sets (e.g. WMAP3) is dominated by uncertainty in the cosmological
parameters; in contrast, the variance in the most recent parameter sets is 
comparable to the variation between fitting functions across the mass range 
considered. This implies that future constraints on cosmology using the HMF 
will be limited by uncertainties in our knowledge of the HMF fitting function 
itself, unless more robust non-parametric means are used.

\begin{figure*}
  \centering
  \includegraphics[width=\textwidth,  trim=3.7cm 0cm 3.7cm 0cm, clip=False]{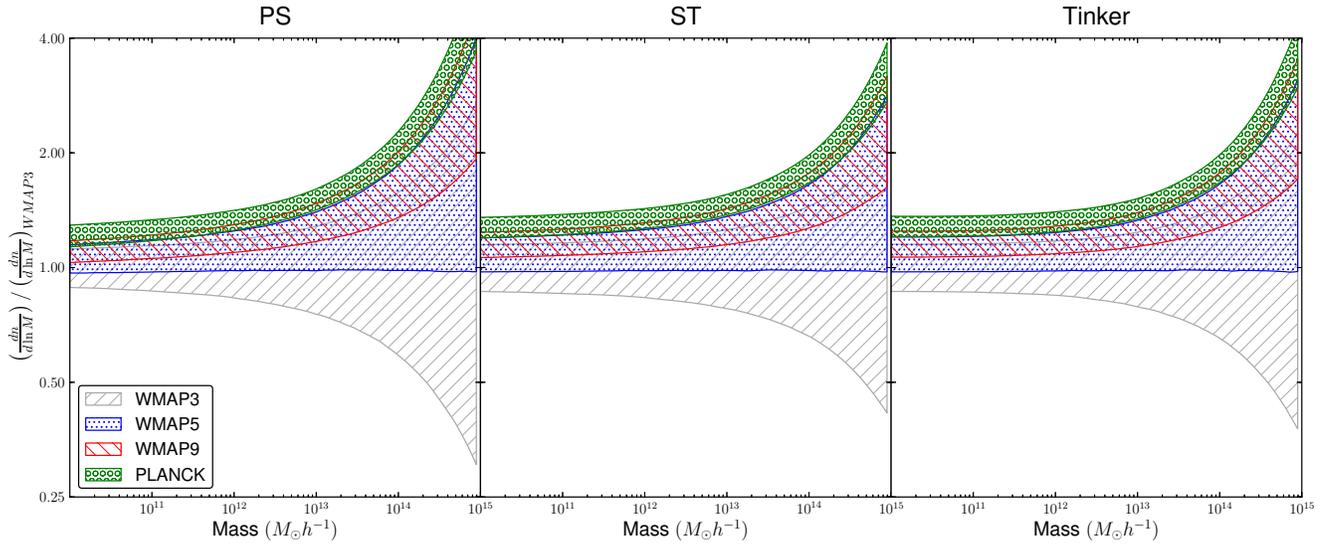}
  \caption{HMFs, with shaded regions indicating 68\% errors, 
    normalized by the median WMAP3 HMF.}
  \label{fig:comparison}
\end{figure*}

More encouraging is the similarity of the several dashed curves in 
Figure \ref{fig:error}, which imply that a change of cosmology (within 
reasonable ranges) does not greatly affect the behaviour of the general 
class of fitting functions\footnote{Even allowing for the 10\% uncertainty 
  arising from non-universality -- see note in \S\ref{sec:method}.}; that is, each 
fitting function displays a similar trend with the change of cosmology. This 
derives from the fitting functions depending on cosmology only through its influence 
on the mass variance (cf. Eq. \ref{eq:hmf}), and justifies our use of a single 
fitting function on which to base a general analysis of the cosmological 
parameter errors. Figure \ref{fig:comparison} confirms this; it shows the 
normalized HMF for several cosmologies for three fitting functions. While there
are noticeable differences between cosmologies, there is consistency of behaviour
between the fitting functions.

\paragraph*{Amplitude of Errors in Standard Cosmologies.}

Figure \ref{fig:error} shows that at low masses, the variation 
asymptotes to a constant value, while at high masses the error rises 
seemingly exponentially. Indeed, for early parameter sets, cluster-mass
haloes had associated uncertainties of almost 70\%. This increase in 
uncertainty at high masses reflects the sensitivity of the abundance of 
clusters to the value of $\sigma_8$, which has been noted 
previously \citep{Vikhlinin2009,Allen2011} and will be discussed further. 
From the latest Planck measurements, we deduce a `rule-of-thumb' of 
$\sim$6\% uncertainty in both the amplitude and slope of the HMF on scales 
less than $10^{13}{\rm M}_{\sun}h^{-1}$, rising to $\sim$20\% for the 
amplitude and $\sim$15\% for the slope at scales of 
$10^{15}{\rm M}_{\sun}h^{-1}$.

\begin{figure}
  \centering
  \includegraphics[width=\linewidth , trim=3cm 0cm 3cm 0cm, clip=True]{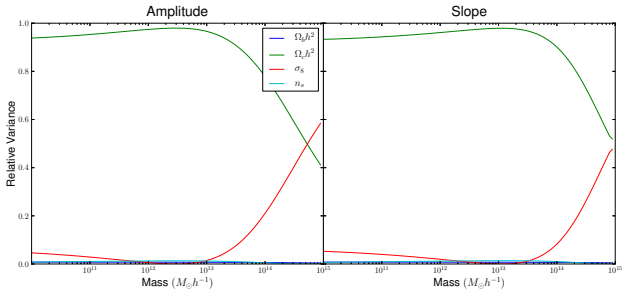}
  \caption{Contribution of single parameters to the overall variance. Each 
    curve is the error of a single parameter as a fraction of the error summed 
    in quadrature of all four parameters. Errors are taken from the Planck cosmology. 
    $\Omega_{\rm c}h^2$ is the dominant term for most of the mass range, with $\sigma_8$ 
    becoming extremely important at high mass.}
  \label{fig:contribution}
\end{figure}

Furthermore, although we observe a decrease in uncertainty over the 
consecutive WMAP and Planck measurements, the rate of decrease is becoming
smaller, especially at high masses. This occurs because of the dominance of
$\sigma_8$, which is the primary contributor to the uncertainty at high masses 
(cf. Figure \ref{fig:contribution}). Therefore, a significantly more 
constrained HMF will depend strongly on the $\sigma_8$ constraint. Maximum likelihood 
techniques applied to observed clusters may be the best way to constrain $\sigma_8$, 
given the strong sensitivity in this regime. Such a constraint could be obtained
from clusters drawn from the XMM XXL X-ray survey, which is designed to identify galaxy 
clusters out to $z \sim 2$ with masses of $M_{\rm vir} \gtrsim 10^{14} {\rm M}_{\odot}$ in 
50 square degrees of sky \citep[cf.][]{Pierre2011}.

 \paragraph*{Consistency of Parameter Sets.}

Figure \ref{fig:comparison} shows the HMF, with error ranges (given by 
the 16$^{th}$ and 84$^{th}$ quantiles) for several parameter sets, 
normalized by the HMF of the mean parameters from WMAP1. This gives a 
visual clue as to the relative amplitude of the HMFs and 
their error, which reveals whether they are statistically consistent.
We see that while Planck results are marginally consistent with WMAP9, 
they do not overlap with WMAP5 over the entire range of masses 
considered. This is driven by the inconsistency of $\Omega_{\rm c}h^2$ between 
these parameter sets. 

\paragraph*{Effects of Single Parameters.}

To better understand the parameters that are most important in the variance 
of the HMF, we marginalise over all but one of the parameters at a time, 
calculating the uncertainty, and plot the fractional uncertainty of 
each parameter with respect to the total uncertainty (summed in 
quadrature) in Figure \ref{fig:contribution}. These results are for the 
Planck parameter set.

The dark matter density parameter $\Omega_{\rm c}h^2$
is the dominant source of uncertainty in both the amplitude and slope of 
the HMF except at high masses. $\Omega_{\rm c}h^2$ has the intuitive effect of 
increasing (decreasing) the amplitude of the HMF as it increases (decreases). 
At high masses, the effect of $\sigma_8$ is even more important than $\Omega_{\rm c}h^2$
because it regulates the amplitude of density perturbations.
Most fitting functions contain an explicit inverse exponential dependence on the 
mass variance, which is linear in $\sigma_8$, so for large masses 
(where the mass variance is very small), the HMF is exponentially sensitive to 
this parameter.

\paragraph*{Application to Non-Standard Dark Matter Models.}

We expect the HMF to be sensitive to the nature of dark matter. Warm Dark 
Matter (WDM) models, in which the abundance of low-mass haloes is suppressed, 
provide a popular alternative to the CDM model and there
has been much interest in them recently \citep[e.g.][]{Smith2011,Schneider2013}. 
The signature of WDM should be evident in the HMF on galaxy-mass scales and
below, whereas most work on the HMF has focussed on group- to cluster-mass scales.
This raises the question -- what is the maximum mass 
scale for a given WDM model which can theoretically differentiate it 
from CDM given the current best cosmological parameters? Or conversely, 
how well constrained must the parameters be to differentiate reasonable 
WDM models at a certain mass? These questions are important for future 
efforts to detect a WDM signal in the HMF, such as, for example, the
GAMA survey \citep{Driver2011,Robotham2011} that aims to measure 
the HMF down to scales of $10^{12}{\rm M}_{\sun}h^{-1}$. To answer 
them, one needs to know the intrinsic uncertainty in the CDM mass 
function, and deduce where the particular WDM model becomes 
significantly different.

\begin{figure}
  \centering
  \includegraphics[width=\linewidth]{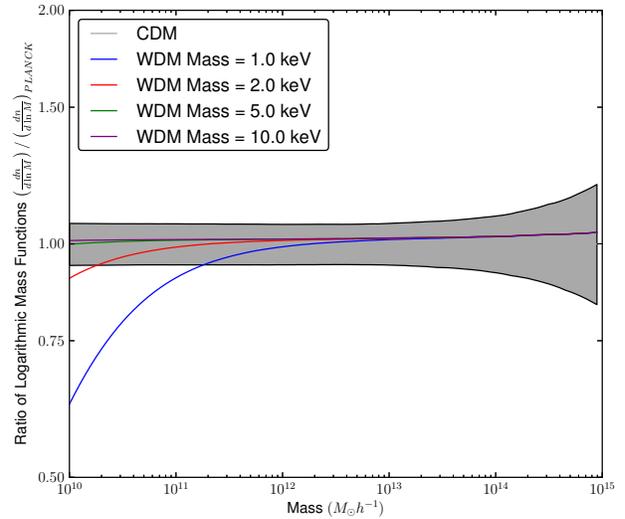}
  \caption{Theoretical HMFs for WDM models with specified 
    particle masses in keV. Overplotted is the error region for the Planck 
    cosmology. This gives an indication of the maximum mass which can be 
    used to differentiate a particular WDM model.}
  \label{fig:wdm}
\end{figure}

Figure \ref{fig:wdm} shows the results of calculating the CDM HMF
for the Planck parameters, with error region shaded, compared
to equivalent WDM HMFs for candidate particle masses between 1 and 10 keV. 
Here we assume the transfer function of \cite{Bode2001} is applied to the 
CDM linear matter power spectrum to produce the WDM linear matter power 
spectrum, required to compute the mass variance $\sigma$ in Eq. \ref{eq:hmf}.
For a plausible WDM particle mass of 1 keV -- in the sense that it is 
consistent with observational limits on the WDM particle mass -- we find that the
maximum mass scale at which the CDM and WDM models are inconsistent is 
$\sim$10$^{11} {\rm M}_{\sun}h^{-1}$. Increasing the particle mass should mimic a
``colder'' dark matter model and reduce the discrepancy with the CDM HMF,
which is what the results show.

By observing that the dominant source of uncertainty at intermediate 
mass scales is $\Omega_{\rm M}$, we can calculate error-ranges on the 
HMF for a given error in this parameter (marginalising over other 
insignificant parameters). By doing so, we may calculate the necessary 
uncertainty required to theoretically distinguish a WDM model with given 
particle mass from the CDM model at a given mass scale. This allows us 
to make a rough estimate of the time it will take to reduce errors in 
$\Omega_{\rm M}$ sufficiently to differentiate reasonable WDM models. We 
can already, in principle, distinguish a 1keV WDM model from the CDM model 
assuming Planck parameters at 10$^{11}{\rm M}_{\sun}h^{-1}$, which is expected 
from figure \ref{fig:wdm}, but it will be 20+ years before we can distinguish
between WDM and CDM on group mass scales ($\sim 10^{13}{\rm M}_{\sun}h^{-1}$).


\paragraph*{Redshift Dependence}
We have repeated our analysis at $z$=1, the redshift 
at which a number of current and future surveys are operating. Results are qualitatively 
similar although we note that 
uncertainties for all base cosmologies are increased by a factor of $\sim$2 across 
the mass range considered. The uncertainty stemming from choice of fitting function 
also increases, and we find that Planck results have smaller uncertainties than the 
fitting functions. This highlights the need for a consistent approach to fitting the HMF, 
and, as argued by \citet{Bhattacharya2011}, raises the question of the suitability of 
simple analytic fits to the mass function for future percent-level analyses.

\section{Summary}
\label{sec:conclusions}

We have examined how our knowledge of the HMF has improved over the last decade with
the availability of increasingly accurate estimates of the cosmological parameters
by the WMAP and Planck satellites. Our analysis reveals the limiting uncertainty in the 
HMF now comes from the variance between fitting functions, with errors intrinsic to 
the parameters being constrained to $\sim$6\% at low masses, and $\sim$20\% at high masses. 
We find that the primary sources of uncertainty in the HMF for a standard flat $\Lambda$CDM 
cosmology are the matter density parameter $\Omega_{\rm M}$ and the normalization $\sigma_8$.
 Of these, $\sigma_8$ is the key parameter which will influence further reductions in the 
uncertainty in
the HMF.

In addition, we have examined prospects for differentiating between the standard CDM
model and its WDM variants using the HMF and found that reasonable models 
only become detectable at scales less than $\sim$10$^{11}{\rm M}_{\sun}h^{-1}$ 
with current parameter uncertainties. We place an upper-limit of 0.8\% 
on the uncertainty in $\Omega_{\rm M}$ before theoretical detection of a WDM 
model of 1keV is possible at $10^{12}{\rm M}_{\sun}h^{-1}$ -- a $\sim$300\% tighter 
constraint than the Planck results. 

We conclude by noting that, while the strongest constraint on the HMF can be made in theory 
using data drawn from the entire halo mass range, this is prohibitively difficult to do in 
practice. Assuming a volume limited sample of haloes, the strongest constraint comes from 
the smallest haloes because of Poisson statistics. However, for an HMF sample constructed 
with a volume limit scaled optimally to match halo mass (i.e. detection of larger haloes 
require larger volumes), there will be a volume disconnect between bins, so the halo mass 
bin to bin variance will dominate HMF shape. This means that we cannot use the entire HMF 
to constrain, say, both dark matter and dark energy models, because the volume overlap 
between observable low mass halos and large clusters is virtually nil. Therefore future 
surveys that seek to use the HMF should target either the most massive haloes -- and tests 
of dark energy -- or the hosts of the lowest mass galaxies -- and tests of dark matter; 
they should not do both.

\section*{Acknowledgements}
ASGR acknowledges support of a UWA postdoctoral research fellowship.
Part of this research was undertaken as part of the 
Survey Simulation Pipeline (SSimPL; {\small ssimpl-universe.tk}). The 
Centre for All-Sky Astrophysics (CAASTRO) is an Australian Research Council Centre of Excellence, 
funded by grant CE11E0090.

We thank the referee for constructive and insightful comments.

\end{document}